\documentclass[aps,twocolumn,superscriptaddress,amsmath,amssymb]{revtex4-1}
 \usepackage{soul}
\usepackage{graphicx}  % needed for figures 
\usepackage{dcolumn}   % needed for some tables
\usepackage{bm}        % for math 
\usepackage{verbatim}   % for math
\usepackage{color} 
\usepackage{ulem} 
\usepackage{physics}
\usepackage{amsmath}
\usepackage{amssymb}
\usepackage[mathlines]{lineno}%
\usepackage{newtxtext,newtxmath}
\usepackage{mathtools}
\DeclareUnicodeCharacter{03B2}{\ensuremath{\beta}}
\let\oldAA\AA
\renewcommand{\AA}{\text{\normalfont\oldAA}}

\usepackage{tikz,xcolor,hyperref}

\definecolor{lime}{HTML}{A6CE39}
\DeclareRobustCommand{\orcidicon}{%
        \begin{tikzpicture}
        \draw[lime, fill=lime] (0,0)
        circle [radius=0.16]
        node[white] {{\fontfamily{qag}\selectfont \tiny ID}};
        \draw[white, fill=white] (-0.0625,0.095)
        circle [radius=0.007];
        \end{tikzpicture}
        \hspace{-2mm}
}
\foreach \x in {A, ..., Z}{%
\expandafter\xdef\csname orcid\x\endcsname{\noexpand\href{https://orcid.org/\csname orcidauthor\x\endcsname}
{\noexpand\orcidicon}}
}

\begin{document}
\title{Carbosilicene and germasilicene: Two $2D$ materials with excellent structural, electronic and optical properties}	

\author{Saeid Amjadian\orcidS} \affiliation{Department of Physics, Iran University of Science and Technology, Narmak, Tehran 16844, Iran}
\author{Mahdi Esmaeilzadeh\orcidM} \affiliation{Department of Physics, Iran University of Science and Technology, Narmak, Tehran 16844, Iran}
\author{Nezhat Pournaghavi\orcidE} \affiliation{Department of Applied Physics, School of Engineering Sciences,
KTH Royal Institute of Technology, AlbaNova University Center, SE-10691 Stockholm, Sweden}

\date{\today}

\begin{abstract}
Using first principle calculations, we study the structural, optical and electronic properties of two-dimensional silicene-like structures of CSi$_7$ (carbosilicene) and GeSi$_7$ (germasilicene) monolayers. We show that both CSi$_7$ and GeSi$_7$ monolayers have different buckling that promises a new way to control the buckling in silicene-like structures. Carbon impurity decreases the silicene buckling, whereas germanium impurity increases it. The CSi$_7$ has semiconducting properties with $0.25\;eV$ indirect band gap, but GeSi$_7$ is a semimetal. Also, under uniaxial tensile strain, the semiconducting properties of CSi$_7$ convert to metallic properties which shows that CSi$_7$ can be used in straintronic devices such as strain sensor and strain switch. There is no important response for GeSi$_7$ under strain. The GeSi$_7$ has higher dielectric constant relative to CSi$_7$, silicene and graphene and it can be used as a 2D-material in high performance capacitors. Calculation of cohesive and formation energies show that CSi$_7$ is more stable than GeSi$_7$. Furthermore, we investigate the optical properties of these new materials and we show that CSi$_7$ and GeSi$_7$ can significantly increase the light absorption of silicene. The obtained results can pave a new route for tuning the electronic and optical properties of silicene like structures for different applications in nanoelectronic devices.
\end{abstract}

\maketitle
\section{Introduction}
Two dimensional (2D) materials like graphene, silicene and germanene are semimetals with zero-gap \cite{w11,cta09}, and their charge carriers are massless fermions\cite{nltzzq12}. Graphene have been studied vastly because of its superior advantages such as mechanical, optical and electronic properties \cite{ajyr19, kjna19, lkk19, lmhlz19, m18, qxz18, rilts19, sjhs19, thxwl20,ytycqw19, zxldxl, pky17,geh17,z16, mwh18}. Different doping are performed in graphene for the new applications such as sulfur-doping for micro-supercapacitors\cite{csqmj19}, nitrogen-doped graphene quantum dots for photovoltaic\cite{hgrpgc19}, silicon nanoparticles embedded in n-doped few-layered graphene for lithium ion batteries\cite{lyzsgc} and implanting germanium into graphene for single-atom catalysis  applications\cite{tmbfbk18}. Theoretical and experimental investigations of graphene-like structures such as silicene and germanene have been vastly carried out \cite{vpqafa,loekve,dsbf12,wqdzkd17,cxhzlt17,ctghhg17}. Silicene and germanene have been grown on Au(111)\cite{cstfmg17}, Ag(111)\cite{jlscl16} and Ir(111)\cite{mwzdwl13} that can encourage researchers to do more study about them. Due to the buckled structure of silicene, it has different physical properties compared to graphene, such as higher surface reactivity\cite{jlscl16}, and a tunable band gap by using an external electric field which is highly favorable in  nanoelectronic devices\cite{nltzzq12,dzf12}. However, the formation of imperfections on the synthesis of silicene is usually inevitable which influences the magnetic and electronic properties of the material\cite{lwtwjl15}. There are some studies about doped atoms such as lithium,  aluminum  and phosphorus in silicene to achieve wide variety of electronic and optical properties\cite{mmt17,dcmj15}. 
Recently simulation and fabrication of 2D silicon-carbon compounds known as siligraphene (Si$_m$C$_n$) have received more attentions due to their extraordinary electronic and optical properties. For example, SiC$_2$ siligraphene which has been experimentally synthesized\cite{lzlxpl15}, is a promising anchoring material for lithium-sulfur batteries\cite{dlghll15}, a promising metal-free catalyst for oxygen reduction reaction\cite{dlghll15}, and a novel donor material in excitonic solar cells\cite{zzw13}. Also, graphitic siligraphene g-SiC$_3$ in the presence of strain can be classified in different electrical phases such as a semimetal or a semiconductor. g-SiC$_3$ has a semimetallic behavior under compression strain up to 8\%, but it becomes a semiconductor with direct band gap (1.62 eV) for 9\% of compression strain and becomes a semiconductor with indirect band gap (1.43 eV) for 10\% of compression strain \cite{dlghll15}. Moreover, g-SiC$_5$  has semimetallic properties and it  can be used as a gas sensor for air pollutant\cite{dwzhl17}. Furthermore, SiC$_7$ siligraphene has a good photovoltaic applications \cite{heakba19} and can be used as a high capacity hydrogen storage material\cite{nhla18}. It shows superior structural, dynamical and thermal stability comparing to other types of siligraphene and it is a novel donor material with extraordinary sunlight absorption\cite{dzfhll16}. The structural and electronic properties of silicene-like SiX and XSi$_3$ (X = B, C, N, Al, P) honeycomb lattices have been investigated\cite{dw13}. Also, the planarity and non-planarity properties for g-SiC$_n$ and g-Si$_n$C (n = 3, 5, and 7) structures have been studied\cite{tllpz19}.

The excellent properties of siligraphene\cite{dzfhll16} motivated us to study CSi$_7$ and GeSi$_7$, in order to find a new approach of silicene buckling and band gap control and to obtain new electronic and optical properties. Here we call CSi$_7$ carbosilicene and GeSi$_7$ germasilicene. We choose carbon and germanium atoms respectively for CSi$_7$ and GeSi$_7$ because these atoms, same as silicon atom, have four valence electrons in their highest energy orbitals. Using density functional theory, we show that both structures are stable but CSi$_7$ is more stable than GeSi$_7$. The carbon atom in CSi$_7$ decreases the buckling, while germanium atom in GeSi$_7$ increases the buckling. It is shown that CSi$_7$ is a semiconductor with 0.24 eV indirect band gap\cite{plgkl20} but GeSi$_7$, similar to silicene, is a semimetal. Also, we investigate the effects of strain and we show that for CSi$_7$, the compressive strain can increase the band gap and the tensile strain can decrease. At sufficient tensile strain (>3.7\%), the band gap of CSi$_7$ becomes zero and thus the semiconducting properties of this material change to metallic properties. As a result, the band gap of CSi$_7$ can be tuned by strain and this material can be used in straintronic devices such as strain sensors and strain switches. For GeSi$_7$, strain does not have any significant effect on it. In contrast, GeSi$_7$ has high dielectric constant and can be used as a 2D material with high dielectric constant in advanced capacitors. Finally, we investigate the optical properties of these materials and we find that the light absorption of both CSi$_7$ and GeSi$_7$ are significantly greater than the light absorption of silicene. Because of high absorption of CSi$_7$ and GeSi$_7$, these materials can be considered as a good candidate for solar cell applications. It is worth to mention that germasilicene, GeSi$_7$, is a new 2D material proposed and studied in this paper, while carbosilicene, CSi$_7$, has been proposed previously as a member of siligraphene but only its band structure has been studied\cite{tllpz19,plgkl19,plgkl20}.
The rest of the paper is organized as follows. In Sec. II, method of calculations is introduced and the results and discussion are given in Sec. III. Section IV contains a summary and conclusion.

\section{Method of calculations}
 Density functional theory (DFT) calculations are performed using the projector-augmented wave pseudopotentials \cite{b94} as implemented in the Quantum-ESPRESSO code\cite{gc09}. To describe the exchange-correlation functional, the generalized gradient approximation (GGA) of Perdew-Bruke-Ernzerhof (PBE) is used\cite{pbe96}. After optimization, the optimum value for the cutoff energy is obtained equal to 80 Ry. Also, Brillouin-zone integrations are performed using Monkhorst-Pack\cite{mp76} and optimum reciprocal meshes of 12×12×1 are considered for calculations. At first, unit cells and atomic positions of both CSi$_7$ and GeSi$_7$ are optimized and then their electronic properties are determined by calculating the density of states and band structure. Moreover, their optical properties are determined by calculating the absorption and the imaginary and real parts of dielectric constant.

\section{Results and discussion}
\subsection{Structural properties}

\begin{figure}[ht!]
\centering
\includegraphics[width=0.98\linewidth,clip=true]{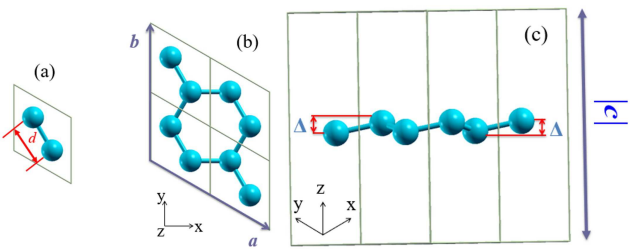}
\caption{(a) Top view of silicene and (b) Si$_8$ unit cells. (c) Side view of Si$_8$ unit cell. 
} 
\label{fig1}
\end{figure}

\begin{figure}[ht!]
\centering
\includegraphics[width=0.8\linewidth,clip=true]{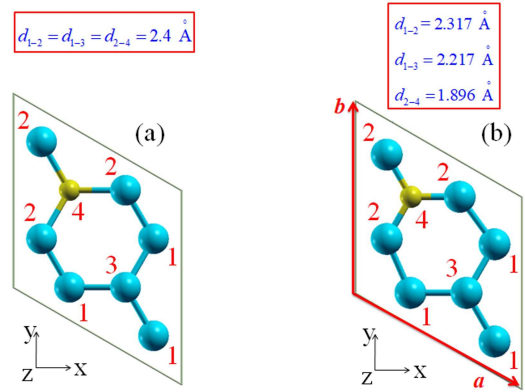}
\caption{Top view of CSi$_7$ unit cell (a) before and (b) after relaxation. Carbon atom is shown by yellow sphere and silicon atoms by blue spheres. 
} 
\label{fig2}
\end{figure}

By increasing silicene unit cell [see Fig.~\ref{fig1} (a)] in x and y direction twice, Si$_8$ has been constructed [see Fig. ~\ref{fig1}(b)] in hexagonal lattice (i.e., $ \alpha=\beta=90^{\circ},\gamma=120^{\circ}$). In physical view, both silicene and Si$_8$ have the same physical properties because by increasing both unit cells, silicene monolayer has been achieved. In this work, Si$_8$ unit cell considered because CSi$_7$ and GeSi$_7$ can be constructed by replacing a silicone atom with a carbon or a germanium atom. After relaxation, the bond length of Si$_8$ was $d=2.4 \;\AA$  [see Fig.~\ref{fig1} (a)] and lattice parameters were $|a|=|b|=7.56 \; \AA$ and $|c|=14.4 \;\AA$ [see Figs.~\ref{fig1}(b) and 1(c)] and buckling parameter $\Delta=0.44 \;\AA$ [see Fig.~\ref{fig1}1 (c)] which has a good agreement with previous works\cite{wwltxh,gzj12,zlyqzw16}. Here c is the distance to make sure that there is no interaction between adjacent layers.                                                                
For carbosilicene, CSi$_7$, unit cell construction, a silicon atom can be replaced with a carbon atom as shown in Fig. ~\ref{fig2}. Because of structural symmetry of CSi$_7$ monolayer (see Fig. ~\ref{fig6}), the position of impurity atom is not important, and our calculations also show the same ground state energy for all the eight possible impurity positions. After relaxation, optimum lattice parameters are obtained as $|a|= |b|=7.49\; \AA$ and $|c|= 12.86 \; \AA$ for CSi$_7$ unit cell. Fig. ~\ref{fig2} shows this structure before and after relaxation. For a more detailed explanation, we labeled atoms in this figure. It is observed that Si-C bond length (i.e., $d_{2-4}=1.896 \; \AA$) is shorter than Si-Si band length (i.e.,  $d_{1-2}=2.317,\; d_{1-3}=2.217 \; \AA$) because of sp$^2$ hybridization. Also, unlike graphene, the hexagonal ring is not a regular hexagon due to the electronegativity difference between C and Si atoms\cite{dzfhll16}.

\begin{figure}[ht!]
\centering
\includegraphics[width=0.98\linewidth,clip=true]{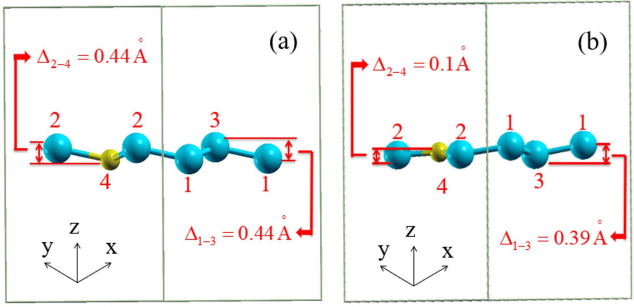}
\caption{Side view of CSi$_7$ unit cell (a) before and (b) after relaxation.
} 
\label{fig3}
\end{figure}

Fig. ~\ref{fig3} shows the side view of CSi$_7$ unit cell. After relaxation, the buckling parameter between atoms 1 and 3 ($\Delta_{1-3}$) is 0.1 $\AA$ whereas this parameter for atoms 2 and 4 ($\Delta_{2-4}$) is 0.39 $\AA$. So, CSi$_7$ has a structure with two different buckling parameters and one can use the carbon atoms to decrease buckling parameter of silicene. Silicene has one buckling and two sublattices\cite{zyy15}, while carbosilicene has two bucklings and thus three sublattices including one for carbon atoms and two others for silicon atoms.

\begin{figure}[ht!]
\centering
\includegraphics[width=0.8\linewidth,clip=true]{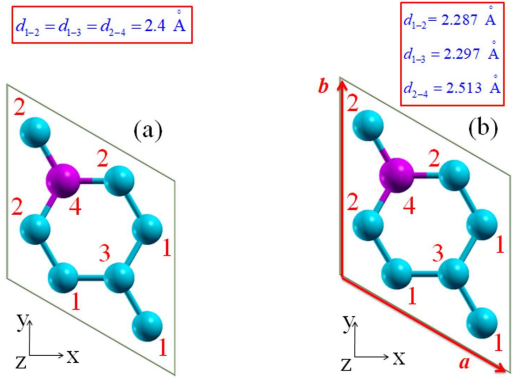}
\caption{ Top view of GeSi$_7$ unit cell (a) before and (b) after relaxation. Here germanium atom is shown by purple color.
} 
\label{fig4}
\end{figure}

If we replace a silicon atom with a germanium atom as shown in Fig.~\ref{fig4}, we could obtain germasilicene, GeSi$_7$, structure. As we can see in this figure, the optimized parameters are $|a|$=$|b|$=7.8$ \AA$, $|c|$=11.98 $\AA$ and the Si-Ge bond length and lattice constants are greater than those of Si-Si. Also, by comparing bond lengths and lattice parameters of GeSi$_7$ and CSi$_7$ structures, it is seen that the bond lengths and lattice parameters of GeSi$_7$ are significantly greater than those of CSi$_7$ which is due to the larger atomic number and thus atomic radius of germanium relative to the carbon\cite{zyihm18}.

\begin{figure}[ht!]
\centering
\includegraphics[width=0.98\linewidth,clip=true]{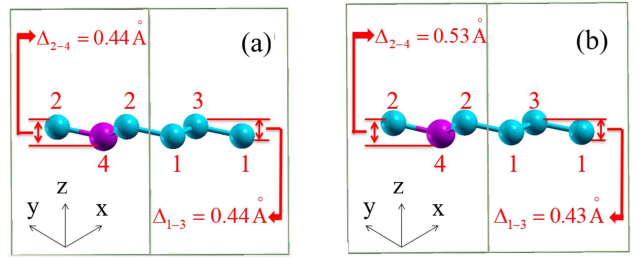}
\caption{ Side view of GeSi$_7$ unit cell (a) before and (b) after relaxation 
} 
\label{fig5}
\end{figure}

\begin{figure}[ht!]
\centering
\includegraphics[width=0.98\linewidth,clip=true]{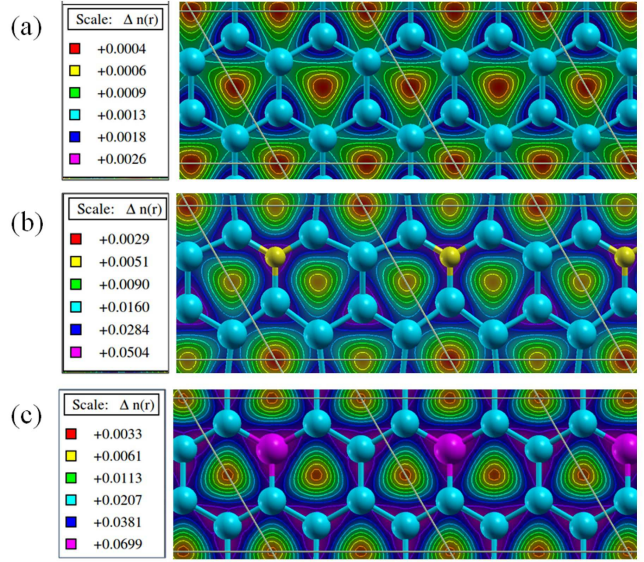}
\caption{ Charge density of (a) silicene, (b) CSi$_7$ and (c) GeSi$_7$ 
} 
\label{fig6}
\end{figure}

The buckling parameters of germasilicene structure are depicted in Fig. ~\ref{fig5}. After relaxation, we find that the value of these parameters are $\Delta_{2-4}=0.53\; \AA$ and $\Delta_{1-3}=0.43 \; \AA$. Therefore, GeSi$_7$ like CSi$_7$ has a structure with two different buckling and the germanium impurity atom increases the buckling of silicene. Bond length values and other structural parameters after relaxation are shown in Table 1.
\begin{table*}[t]
  \centering
    \begin{tabular}{|c|c|c|c|c|c|c|c|c|} \hline
        &$|a|=|b|$&	$|c|$	&$d_{1-2}$	&$d_{2-4}$&	$d_{1-3}$	&$\Delta_{2-4}$&	$\Delta_{1-3}$&	$\Delta_d$  \\
        \hline
        Si$_8$	& 7.65 & 	14.4 &	2.4 &	2.4	& 2.4 &	0.44 &	0.44 &	0 \\
        \hline
        CSi$_7$ &	7.49 &	12.86 &	2.317 &	1.896 &	2.217 &	0.1	& 0.39 &	0.29\\
        \hline
        GeSi$_7$ &	7.8 &	11.98 &	2.287 &	2.34 &	2.297 &	0.53 &	0.43 &	0.1\\
        \hline
        
    \end{tabular}
    \caption{Optimum lattice parameters $|a|$, $|b|$ and $|c|$, bond lengths $d_{1-2}$, $d_{2-4}$ and 
$d_{1-3}$ and buckling parameters $\Delta_{2-4}$,  $\Delta_{1-3}$ and  $\Delta_d$. All values are in Angstrom.
}
    \label{tab1}
\end{table*}

We now introduce a new parameter for buckling as 
\begin{equation}
    \Delta_d=|\Delta_{2-4}-\Delta_{1-3}|
    \label{eq1}
\end{equation}
which shows the difference between  two buckling parameters. Value of $\Delta_d$ for CSi$_7$ (i.e., 0.29 $\AA$) is greater than that for GeSi$_7$ (i.e., 0.062 $\AA$) which means the carbon impurity atom has a greater impact than germanium on silicene buckling. This effect could be explained based on electronegativity difference\cite{drsbf13}. The electronegativity by Pauling scale is 2.55 \cite{ipl09,zhhkl20}, 1.9 \cite{gperdd20} and 2.01 \cite{mgyz19} for carbon, silicon, and germanium respectively. Therefore, electronegativity difference is 0.65 for CSi$_7$ and 0.11 for GeSi$_7$ which show that CSi$_7$ has a greater electronegativity difference which leads to the  in-plane hybridized bondings and reduces the buckling in comparison to the other cases.

Fig. ~\ref{fig6} shows the charge density of a monolayer of CSi$_7$ and GeSi$_7$. The charge density of a monolayer of Si is also shown in this figure for comparison [see Fig. ~\ref{fig6}(a)]. The high charge density around the carbon and germanium impurity atoms [see Figs. ~\ref{fig6}(b) and 6(c)] shows charge transfer from silicon atoms to impurity atoms. Also, the electron aggregation around impurity atoms indicates ionic- covalent bonds in CSi$_7$  and GeSi$_7$ structures because of electronegativity difference.

Now, we calculate the cohessive and formation energies for these structures. The cohessive energy is -4.81 eV/atom and -4.32 eV/atom for CSi$_7$ and GeSi$_7$, respectively. The negative value of cohecive energy for CSi$_7$ and GeSi$_7$ means that these structures will not be decomposed into their atoms. The more negative cohesive energy, the more stable structure, so CSi$_7$ is more stable than GeSi$_7$. Also, the caculated cohesive energy for silicene is -4.89 eV/atom which is in good agreement with previous studies \cite{gperdd20,mgyz19} and shows CSi$_7$ has a stable structure with cohessive energy very close to silicene.
Our calculations show the formation energy for CSi$_7$ and GeSi$_7$ structures are +0.16 eV/atom and -0.005 eV/atom, respectively. So, the formation of CSi$_7$ (GeSi$_7$) from their constituents is endothermic (exothermic) because of the positive (negative) value of formation energy. On the other hand, positive formation energy for CSi$_7$ represents a high stability of this structure, while the negative or nearly zero value for GeSi$_7$ is attributed mostly to the high reactivity related to silicene\cite{dw13}.

\subsection{Electronic properties}
\begin{figure}[ht!]
\centering
\includegraphics[width=0.7\linewidth,clip=true]{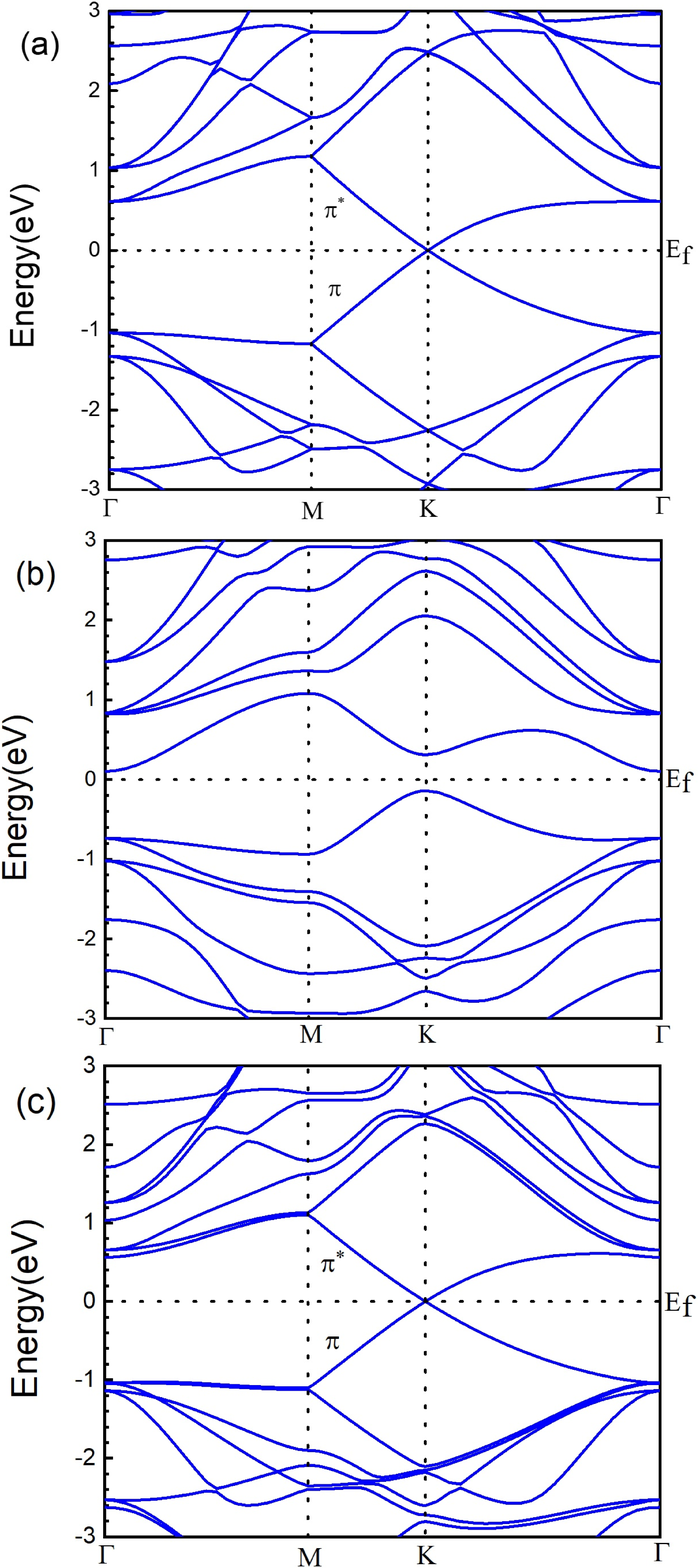}
\caption{ Band structure of (a) silicene, (b) CSi$_7$ and (c) GeSi$_7$. 
} 
\label{fig7}
\end{figure}

To investigate electronic properties of CSi$_7$ and GeSi$_7$, at first, we compare band structure of silicene, CSi$_7$ and GeSi$_7$ monolayers and we show the results in Fig. ~\ref{fig7}. As we can see in this figure, like graphene and silicene, GeSi$_7$ is semi-metal (or zero-gap semiconductor) with Dirac cone in point $K$. This is because the $\pi$ and $\pi^*$ bands cross linearly at the Fermi energy $E_F$. These band structures indicate that the charge career in silicene and GeSi$_7$ behave like massless Dirac fermions\cite{zlyqzw16}. In contrast with GeSi$_7$, CSi$_7$ is a semiconductor with indirect band gap. The value of its indirect band gap is 0.24 eV in $K-\Gamma$ direction which significantly less than its direct band gap value (i.e., 0.5 eV in $K-K$ direction).

\begin{figure}[ht!]
\centering
\includegraphics[width=0.9\linewidth,clip=true]{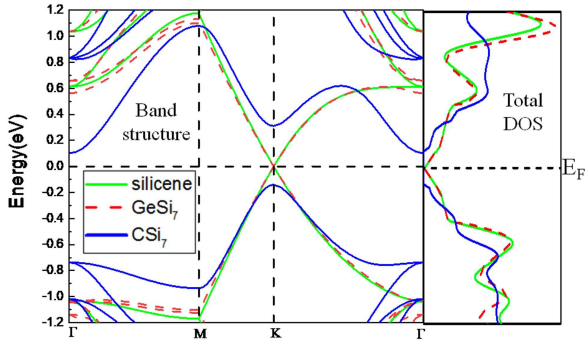}
\caption{ Enlarged band structure and total DOS of silicene, CSi$_7$ and GeSi$_7$. 
} 
\label{fig8}
\end{figure}

For a better comparison, an enlarged band structure of silicene, CSi$_7$ and GeSi$_7$ are shown in Fig. ~\ref{fig8}. It is seen that, in point $K$, silicene and GeSi$_7$ have similar band structures with zero band gap, whereas CSi$_7$ has a band gap. In Dirac cone of graphene and silicene, $\pi$ and $\pi^*$ bands are made from the same atoms\cite{dw13} but these bonds in GeSi$_7$ are made from two different atoms. To determine the Fermi velocity, $v_F$, the graphs for silicene and GeSi$_7$ must be fitted linearly near the Fermi level by using equation $E_{k+K}=\gamma k$. Then the Fermi velocity is given by $v_F=\gamma/ \hbar$. Our calculations show that $v_F$ is $5\cross10^5$  m/s for silicene (which shows a good agreement with previous works\cite{dw13,wd13})  and $4.8\cross10^5$  m/s for GeSi$_7$. A comparison between Fermi velocity in silicene and GeSi$_7$ indicates that Ge atoms in GeSi$_7$ do not have a significant effect on Fermi velocity. The total density of states (DOS) is also shown in Fig. ~\ref{fig8}. It is observed that the total DOS has a good agreement with the band structure.

\begin{figure}[ht!]
\centering
\includegraphics[width=.98\linewidth,clip=true]{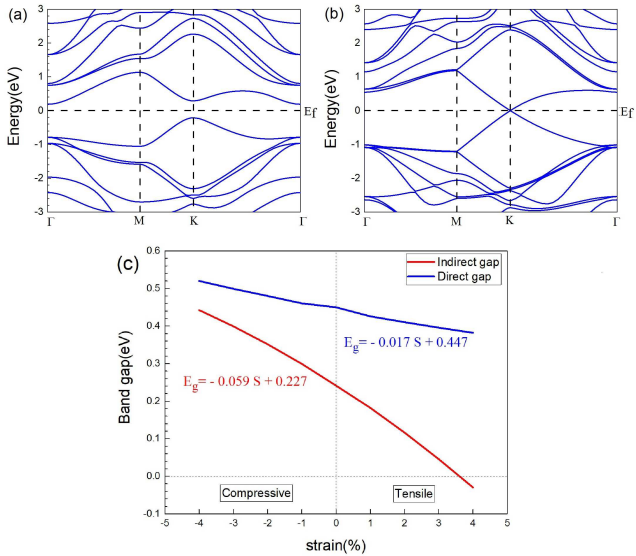}
\caption{ Band structure of (a) CSi$_7$ and (b) GeSi$_7$ under compressive strain with value -3$\%$. (c) Energy gap variation of CSi$_7$ versus both compressive and tensile strains.
} 
\label{fig9}
\end{figure}

We now investigate the effect of strain on the band structure of CSi$_7$ and GeSi$_7$ and the results are shown in Fig. ~\ref{fig9}. As we can see in Figs. ~\ref{fig9}(a) and ~\ref{fig9}(b), compressive strain has important effects on band structure of CSi$_7$ but it has no significant effect on GeSi$_7$ [compare these figures with Figs. ~\ref{fig7}(b) and ~\ref{fig7}(c)]. In the presence of compressive strain for CSi$_7$, both direct and indirect band gaps increase, respectively from 0.5 eV and 0.24 eV to 0.52 eV and 0.44 eV. But for GeSi$_7$, the zero-band gap remains unchanged and compressive strain cannot open any band gap. Fig. ~\ref{fig9}(c) shows the direct and indirect band gap variations of CSi$_7$ versus the both compressive and tensile strains. It is observed that both direct and indirect band gaps increase with increasing the compressive strain, while they decrease with increasing the tensile strain. The variation of band gaps versus strain S is nearly linear and could be formulated by $E_g=-0.017S+0.447$ for direct band gap and $E_g=-0.059 S+0.227$ for indirect one. Under strain and without strain, the direct band gap has significantly larger values relative to indirect band gap, thus it has no important effect on electronic transport properties in CSi$_7$. In contrast with GeSi$_7$, the strain is an important factor for tuning of band gap in CSi$_7$. For example, when the tensile strain increases above   the band gap of CSi$_7$ disappears and this 2D material becomes a metal [see Fig. ~\ref{fig9}(c)]. This property of CSi$_7$ is important in straintronic devices such as strain switches and strain sensors.

\subsection{Optical properties}
\begin{figure}[ht!]
\centering
\includegraphics[width=.7\linewidth,,clip=true]{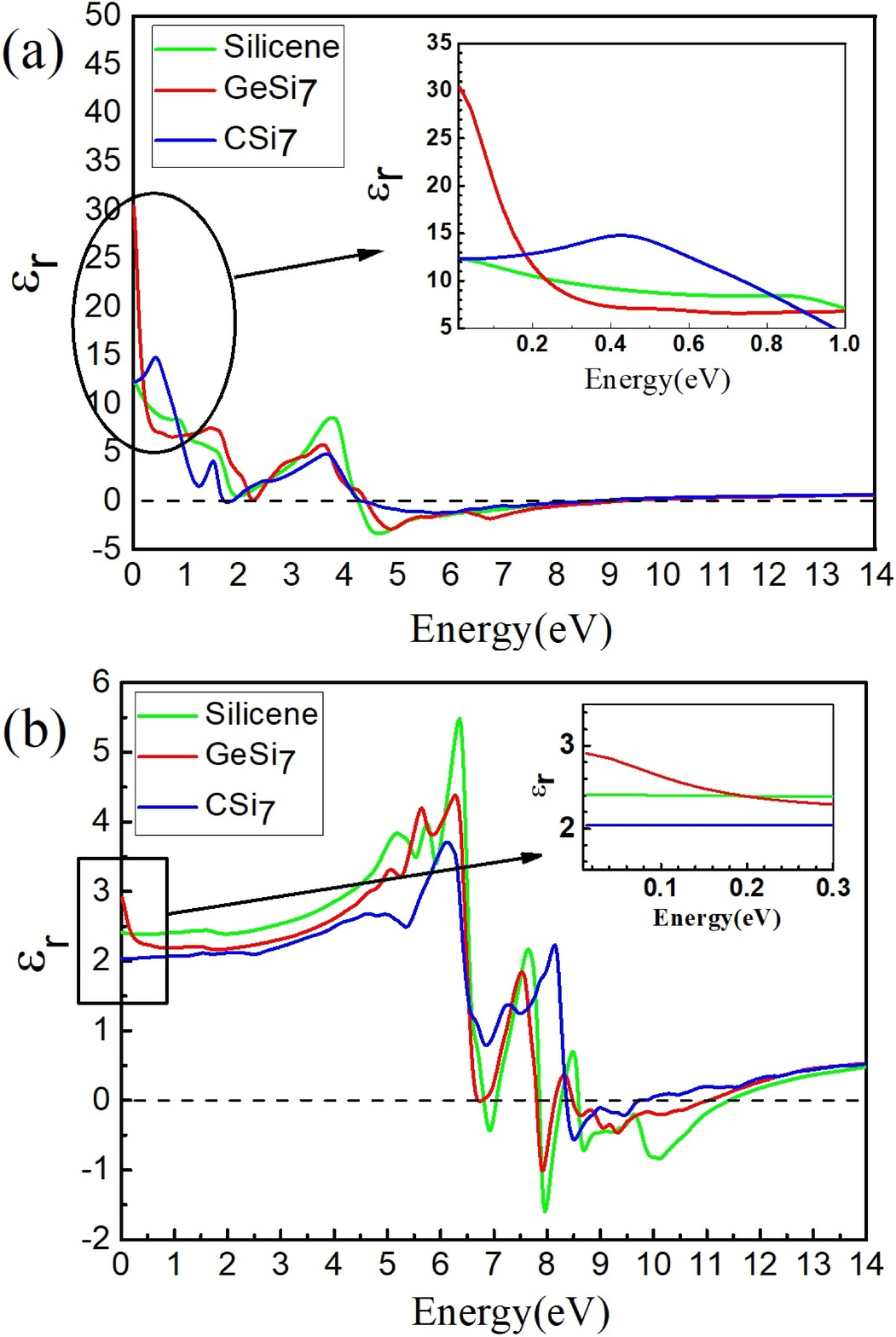}
\caption{ Comparison of real part of dielectric function for CSi$_7$ and GeSi$_7$ (a) in x direction and (b) in z direction. The graphs of silicene are also shown in this figure for comparison.
} 
\label{fig10}
\end{figure}

The complex dielectric function $\epsilon=\epsilon_r+\epsilon_i$ can be calculated for both polarizations of light: (i) parallel (x direction) and (ii) perpendicular (z direction), where $\epsilon_r$ is the real part and $\epsilon_i$ is the imaginary part of the dielectric function. This function is an important parameter for calculation of optical properties of matters. For instance, the real and imaginary parts of refractive index (i.e., $n=n_r+n_i$) can be written as\cite{w}

\begin{equation}
    n_r=\sqrt{\frac{(\epsilon_r^2+\epsilon_i^2)^{1/2}+\epsilon_r}{2}}
    \label{eq2}
\end{equation}
and
\begin{equation}
      n_i=\sqrt{\frac{(\epsilon_r^2+\epsilon_i^2)^{1/2}-\epsilon_r}{2}}
      \label{eq3}
\end{equation}
respectively. The absorption coefficient $\alpha$ is given by\cite{w}
\begin{equation}
    \alpha=\frac{2\omega n_i}{C}
    \label{eq4}
\end{equation}
where C is the speed of light in vacuum. The real parts of dielectric function of CSi$_7$, GeSi$_7$ and silicene are depicted in Fig. ~\ref{fig10} for x and z directions. This figure shows that $\epsilon _r$ in both directions are inhomogeneous because the graphs of $\epsilon_r$ are not similar for the two directions. The root of real part (where $\epsilon_r=0$) represents the plasma energy (frequency) which for these materials it locates at $4.3\; eV \;(1.04\;PHz)$ for x-direction.  It can be seen from Figs. ~\ref{fig10}(a) and ~\ref{fig10}(b) that the values of static dielectric constant (the value of dielectric function real part at zero frequency or zero energy) in the x-direction are 12.3 for silicene and CSi$_7$ and 30 for GeSi$_7$, and in the z-direction are 2.4, 2 and 2.9 for silicene, CSi$_7$ and GeSi$_7$ respectively. Thus, for both directions GeSi$_7$ has the biggest static dielectric constant. Also, the static dielectric constant of GeSi$_7$ is significantly greater than graphene (1.25 for z-direction and 7.6 for x-direction\cite{rdj14}). According to the energy density equation of capacitors (i.e., $u=\epsilon\epsilon_0 E^2/2$), by increasing dielectric constant $\epsilon$, the energy density u  increases. Here, E in the electric field inside the capacitor. So, materials with high dielectric constant have attracted a lot of attentions because of their potential applications in transistor gate, non-volatile ferroelectric memory and integral capacitors\cite{tic06}. Among the 2D-materials, graphene has been used for electrochemical capacitors\cite{cls13} and supercapacitors\cite{vrsgr08}. Since GeSi$_7$ has a high dielectric constant, it can be used as a 2D-material with high performance dielectric in advanced capacitors. 

\begin{figure}[ht!]
\centering
\includegraphics[width=0.75\linewidth,clip=true]{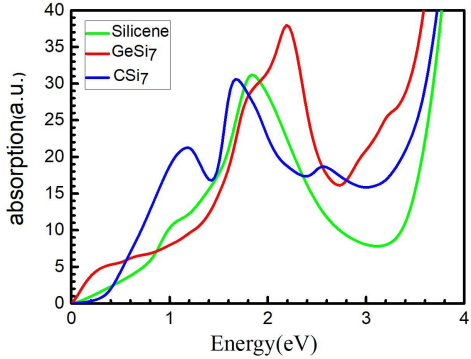}
\caption{ Absorption coefficient for silicene, CSi$_7$ and GeSi$_7$.
} 
\label{fig11}
\end{figure}

Fig. ~\ref{fig11} shows absorption coefficient $\alpha$ for CSi$_7$ and GeSi$_7$. Absorption coefficient for silicene is also shown in this figure for comparison. The absorption coefficient shown in this figure for silicene is in agreement with previous works\cite{hkb18,cjlmty16}. There are two peaks for CSi$_7$: one locates in 1.18 eV (infrared region) and the other in 1.6 eV (visible region). The peak for silicene (at 1.83 eV) locates in visible region (1.8-3.1 eV). So, carbon atom increases and shifts the edge of absorption from the visible region to infrared region because it breaks the symmetry of silicene structure and it opens a narrow energy band gap in silicene band structure. For GeSi$_7$ there is an absorption peak in visible region (at 2.16 eV). Also, the peak height of GeSi$_7$ is larger than that of silicene and CSi$_7$. The sun light spectrum includes different wavelengths and absorption of each part has a special application. For example, ultraviolet-visible region absorption spectrophotometry and its analysis  are used in pharmaceutical analysis, clinical chemistry, environmental analysis and inorganic analysis\cite{rop88}. Also near infrared ($\lambda$= 800 to 1100 nm or E = 1.55 eV to 1.13 eV) and infrared ($\lambda$ > 1100 nm or E < 1.13eV) regions are used for solar cells\cite{wrmss,sgzlgp20}, latent fingerprint development\cite{bsckm19}, brain stimulation and imaging\cite{cwcgy20}, photothermal therapy\cite{hhwl19}, photocatalysis\cite{qzzwz10}  and photobiomodulation\cite{whwlh17}.

\begin{figure}[ht!]
\centering
\includegraphics[width=0.9\linewidth,clip=true]{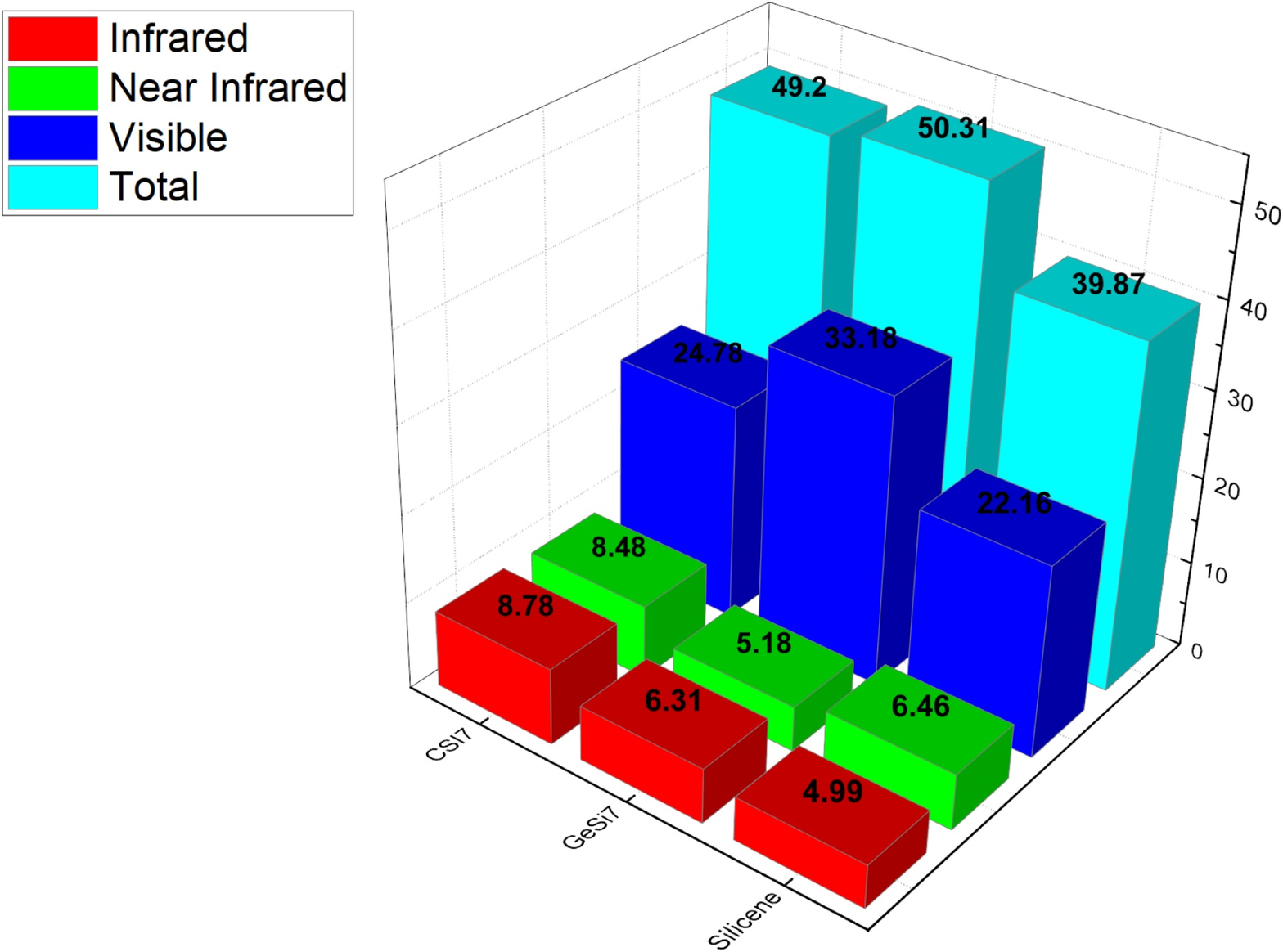}
\caption{ Areas under the absorption curve for silicene, CSi$_7$ and GeSi$_7$ in infrared, near infrared and visible spectrum regions.
} 
\label{fig12}
\end{figure}

On the other hand, sunlight radiation received by earth is comprising  5$\%$ u$\%$ltraviolet, 45$\%$ infrared and 50$\%$ visible \cite{hs11}. So, we investigate area under the absorption curve of CSi$_7$ and GeSi$_7$ in visible (from 1.8 to 3.1 eV), near infrared (from 1.13 to 1.55 eV) and infrared (<1.13 eV). Fig. ~\ref{fig12} shows this area for silicene, CSi$_7$ and GeSi$_7$ in infrared, near infrared and visible spectrum regions. As we can see in this figure, the absorption of CSi$_7$ for all three spectrum regions and total absorption are significantly greater than those of silicene. The absorption of GeSi$_7$ is greater than that of silicene in infrared and visible regions and it is smaller in near infrared region, but the total absorption of GeSi$_7$ is significantly greater than the total absorption of silicene. For comparison, we also calculate the absorption coefficient in infrared region for siligraphene SiC$_7$ , a new material studied recently\cite{dzfhll16}. The absorption for siligraphene in infrared region is equal to 2.7 which shows that CSi$_7$ with absorption 8.78 and GeSi$_7$ with absorption 6.31 have more than two times greater absorption relative to siligraphene in infrared region.

\subsection{Summary and conclusion}
We studied the structural, electronic and optical properties of CSi$_7$ and GeSi$_7$ structures using density functional theory within Quantum Espresso code. We showed that the carbon atom in CSi$_7$ decreases the buckling, whereas germanium atom in GeSi$_7$ increases the buckling which promises a new way to control the buckling in silicene-like structures. Both structures are stable but CSi$_7$ is more stable than GeSi$_7$. Band structure and DOS plots show CSi$_7$ is a semiconductor with 0.24 eV indirect band gap but GeSi$_7$, similar to silicene, is a semimetal. Strain does not have any significant effect on GeSi$_7$, but for CSi$_7$, the compressive strain can increase the band gap and tensile strain can decrease it. At sufficient tensile strain ($> 3.7 \%$), the band gap becomes zero or negative and thus the semiconducting properties of CSi$_7$ change to metallic properties. As a result, the band gap of CSi$_7$ could be changed and controlled by strain and this material can be used in straintronic devices such as strain sensor and strain switch.  Furthermore, we investigated the optical properties of CSi$_7$ and GeSi$_7$ such as static dielectric constant and light absorption. The GeSi$_7$ has high dielectric constant relative to CSi$_7$, silicene and graphene and can be used as a 2D-material with high performance dielectric in advanced capacitors. The light absorption of CSi$_7$ for near infrared, infrared and visible regions and its total absorption are significantly greater than those of silicene. The absorption of GeSi$_7$ is greater than that of silicene in infrared and visible regions and it is smaller in near infrared region, but the total absorption of GeSi$_7$ is significantly greater than the total absorption of silicene. Because of high absorption of CSi$_7$ and GeSi$_7$, these materials can be considered as proper candidates to solar cell applications.
\bibliography{SiC}
\end{document}